\documentclass[11pt,a4paper]{article}
\pdfoutput=1
\usepackage{jcappub}
\usepackage{graphicx}
\usepackage{amsmath}
\usepackage{epsfig,multicol,bbm}
\usepackage{subcaption}

\newcommand{\mpc}{$h^{-1}$Mpc}
\newcommand{\vmin}{$\mathbf{v}_\text{min}$}
\newcommand{\vb}{$\mathbf{v}_\text{bulk}$}

\begin{document}

\title{Frames of most uniform Hubble flow}

\abstract{It has been observed \citep{Wilthvar,Wilthvar2} that the locally measured Hubble parameter converges quickest to the background value and the dipole structure of the velocity field is smallest in the reference frame of the Local Group of galaxies. We study the statistical properties of Lorentz boosts with respect to the Cosmic Microwave Background frame which make the Hubble flow look most uniform around a particular observer. We use a very large N-Body simulation to extract the dependence of the boost velocities on the local environment such as underdensities, overdensities, and bulk flows.  We find that the observation \citep{Wilthvar,Wilthvar2} is not unexpected if we are located in an underdensity, which is indeed the case for our position in the universe. The amplitude of the measured boost velocity for our location is consistent with the expectation in the standard cosmology. %There is however an anomaly in the ... which requires further investigation.
}

\author[a]{David Kraljic}
\author[a,b]{\& Subir Sarkar}

\affiliation[a]{Rudolf Peierls Centre for Theoretical Physics,
  University of Oxford,\\ 1 Keble Road, Oxford, OX1 3NP, United
  Kingdom}
\affiliation[b]{Niels Bohr International Academy, University of
  Copenhagen,\\ Blegdamsvej 17, 2100 Copenhagen, Denmark}
\emailAdd{David.Kraljic@physics.ox.ac.uk}\emailAdd{Subir.Sarkar@physics.ox.ac.uk}
\keywords{cosmic flows, cosmological simulations}
\arxivnumber{}
\maketitle
\flushbottom

\section{Introduction}

The standard Lambda Cold Dark Matter ($\Lambda$CDM) cosmological model is based on an assumed background Friedmann-Lema\^itre-Robertson-Walker (FLRW) geometry that is isotropic and homogeneous. The structure in the universe is modelled as statistically isotropic and homogeneous, initially Gaussian distributed perturbations to this background. This would result in some scatter in the Hubble diagram due to local `peculiar' (non-Hubble) velocities. The cosmic rest frame is defined by comoving observers in the FLRW background and the Cosmic Microwave Background (CMB) frame is conventionally taken to correspond to this `standard of rest' in which both the leading order linear Hubble law and peculiar velocities are defined. The observed dipole pattern in the CMB is then interpreted as due to our peculiar velocity with respect to the cosmic rest frame. 

For an ensemble of observers, the frame where the velocity flow around an average observer is most uniform corresponds to the CMB frame. However, for a particular observer, the most uniform flow may well be in a frame boosted with respect to the CMB. In this paper we are primarily concerned with the reference frame wherein the Hubble parameter in successive radial shells converges most quickly to the background value. We call this the `minimum Hubble variation frame' or frame with most uniform Hubble flow. %Given only local measurements of the velocity field and the assumption that the observer is not in a special position (a.k.a. the `Cosmological Principle'), a non-CMB frame may well be chosen as the cosmic rest frame.

The authors of \citep{Wilthvar,Wilthvar2} looked for a different standard of rest based on the uniformity of the Hubble flow. They found that the Hubble parameter averaged in radial shells converges quickest to its background value in a frame that is boosted with respect to the CMB frame and corresponds roughly to the frame of the Local Group (LG) of galaxies. Moreover, the dipole structure of the velocity flow persists in the CMB frame at large distances, contrary to expectation, but is smaller after boosting to the LG frame. Both of these observations make the local universe as seen in the boosted LG frame closer to the usual ($\Lambda$CDM) expectation. It was suggested \citep{Wilthvar2} that the velocity of the minimum Hubble variation frame should correspond roughly to the group velocity of the `finite infinity' region \citep{EllisFI,WiltFI, WiltFI2}.  Ref.\citep{Bolejkodiff} continued the study of the Hubble flow anisotropy with emphasis on the non-kinematic differential expansion of space as the origin of (at least a part) of the CMB dipole.  Note that this general relativistic effect where the Hubble parameter is both a function of space and time is \emph{not} captured in N-Body simulations where by construction there is a single background expansion rate.

These studies \citep{Wilthvar,Wilthvar2,Bolejkodiff} were in part motivated by some analyses of the local bulk flow of galaxies (as measured in the CMB frame) which show a lack of convergence to the CMB frame even beyond $\sim100$ Mpc \citep{lauerbulk,hudbulk2,kashlinskybulk2,watkinsbulk,lavabulk,kashlinskybulk,feldmanbulk,colinbulk,6dfbulk,2012arXiv1210.0625G,macbulk,watkinsbulk2}, as would be the na\"ive expectation if the universe is indeed homogeneous on larger scales as is inferred from galaxy counts in the SDSS \cite{2005ApJ...624...54H} and WiggleZ \cite{2012MNRAS.425..116S} surveys. Other authors have used different methods (and data) to argue however that observed bulk flows are consistent with $\Lambda$CDM \citep{nusserbulk,daibulk,turnbullbulk,mabulk,mabulk2,hongbulk,applbulk,hutebulk,scrimbulk,feinbulk}. The situation is rather confusing e.g. Ref.\citep{colinbulk} showed using 165 SNe~Ia with redshift $z \lesssim 0.1$ in the Union~2 catalogue that there is an anomalously high and apparently constant bulk flow of $\sim250$~km/s extending out to the Shapely supercluster at $z \simeq 0.06$ ($\sim260$~Mpc). This was  confirmed using 117 new SNe~Ia from the Nearby Supernova Factory survey and it was shown that the flow extends out beyond Shapely, nevertheless these authors concluded that their finding is in agreement with $\Lambda$CDM \citep{feinbulk}. SNe~Ia catalogues do have rather incomplete distributions on the sky which can bias the result, however the discrepancy with the standard expectation has been \emph{confirmed} by analysis of the 6dF galaxy redshift survey which is the largest, most homogeneously derived peculiar velocity sample to date \citep{2012arXiv1210.0625G}.

The variance of the Hubble parameter can be calculated in linear perturbation theory (e.g. Ref.\citep{wanglin}) or using N-Body simulations (e.g. Ref.\citep{HubbleNBody}). In the latter study the dependence of the local Hubble parameter on the observer's environment (i.e. halos versus voids) was explored and it was found, as expected, that in voids the local value of the Hubble parameter is biased towards higher values while the opposite is true for observers in halos. Given the evidence that we are located in a large under-dense region \citep{hbubble,hbubble2,underdensity,2014MNRAS.437.2146W} this is particularly relevant for explaining the tension between the locally measured value of $H_0$ and the (smaller) value inferred from fitting CMB data (see e.g. Refs.\citep{marrahubble,bendayanhubble}). Another approach is to reconstruct the local universe in N-Body simulations e.g. Ref.\citep{hesshubble} finds that our particular position does bias the locally measured Hubble parameter upwards by about 2\%. However others conclude that the expected variance is far too small to explain the current discrepancy between local and global probes of $H_0$ \citep{2014JCAP...10..028O,2016JCAP...02..001O}. In Ref.\citep{Liuhubble} it is argued that the Hubble parameter can be measured more precisely if observations are restricted to only the zones around critical points of the velocity field, in contrast with the usual approach of increasing statistics by averaging indiscriminately over large datasets.

In this paper we study the statistical properties of boosted frames in which the spherically averaged Hubble flow looks most uniform. We are specially interested whether such a boost particular to our position (as measured in Ref.\citep{Wilthvar2}) is consistent with the $\Lambda$CDM expectation. We also look at the dipolar structure of the Hubble flow and its role in determining the frames of minimum Hubble variation. We re-derive the expression for the systematic offset of Hubble parameters between different reference frames paying attention to the dipolar structure. Our expression agrees with the measured offset between the CMB and LG frames and resolves much of the discrepancy found in previous studies \citep{Wilthvar,Wilthvar2}.

We use one of the biggest N-Body simulations of $\Lambda$CDM to date \citep{darksky} having a volume of (8\,$h^{-1}$Gpc)$^3$ with the observers placed randomly in large-scale under- and over-densities. We find that the expectations derived from the simulation are consistent with the measurements for our particular position \citep{Wilthvar,Wilthvar2}. Specifically, for an observer in an under-density, a boost simultaneously makes the spherically averaged Hubble parameter converge quicker to the background value and reduces the dipole structure. We also find that the boost velocity of the frame that minimises the Hubble variation for our location is \emph{consistent} with the distribution extracted from the $\Lambda$CDM simulation and is indeed correlated with the group velocity of the `finite infinity' region as has been suggested earlier \citep{Wilthvar2}.

In \S~\ref{sec:sims} we describe the N-Body simulation and the halo finder used in this paper. \S~\ref{sec:frames} summarises the methods and the theoretical concepts used in this paper. Our results are presented and discussed in \S~\ref{sec:results}, and we conclude in \S~\ref{sec:conclusion}.

\section{Simulations and data}
\label{sec:sims}

We use the largest N-Body simulation of the Dark Sky (DS) Simulations Early Data Release \citep{darksky}. It is a dark matter only simulation using $10240^3 \approx 10^{12}$ particles in a volume of $(8000$\,\mpc$)^3$. The cosmology is $\Lambda$CDM with the following parameters: ($\Omega_\text{m}=0.295,\Omega_\text{b}=0.0468,\Omega_\Lambda=0.705,n_\text{s}=0.969, h=0.688, \sigma_8=0.835$). The halo catalogue used in this paper is obtained using \textsc{rockstar} \citep{rockstar}, a phase-space based algorithm at $z=0$. We use the halos from the simulation as discrete tracers from which the velocity field is obtained. To make computation more efficient, we reduce the number of halos by imposing a mass cut ($M_\text{vir} >10^{12}M_{\odot}/h$). The resulting catalogue contains $2.3 \times 10^9 $ DM halos with a mean number density of $4.6 \times 10^{-3}$ halos/(\mpc)$^3$. Equivalently, the mean halo-halo distance is approximately 6\,\mpc. This number density is about twice that of the \textsc{composite} catalogue \citep{feldmanbulk,watkinsbulk} used in Ref.\citep{Wilthvar,Wilthvar2} and comparable to the \textit{Cosmicflows-2.1} \citep{cosmicflows} catalogue used in Ref.\citep{Wilthvar2}. We use both of these in \S~\ref{sec:results}. Although \textit{Cosmicflows-2.1} is affected by Malmquist bias as discussed in Refs.\citep{Wilthvar2,hofbias} we do not attempt to correct for this as we are interested in calculating quantities where this bias cancels.

\section{Frames with minimal Hubble flow variation}
\label{sec:frames}

In this section we study frames of observers where the spherically averaged Hubble flow converges the quickest to the background value. In an universe with structure there is a specific reference frame at each position where the expansion looks most uniform and we can determine its velocity $\mathbf{v}_\text{min}$ relative to the cosmic rest frame. If the universe has no preferred direction, the unique frame where the variation of the Hubble flow is minimised for an ensemble of observers corresponds, by symmetry, to the cosmic rest frame (aka the CMB frame) i.e. $\langle \mathbf{v}_\text{min} \rangle=\mathbf{0}$, where the brackets correspond to the ensemble average. Therefore, we expect a distribution of velocities $\mathbf{v}_\text{min}$ for an ensemble of observers in a $\Lambda$CDM universe and consequently a non-zero expectation for the amplitude of \vmin\ for any particular observer. This directly parallels the consideration of bulk velocities, insofar as $\langle \mathbf{v}_\text{bulk} \rangle=\mathbf{0}$ for any region of space but $\langle \vert \mathbf{v}_\text{bulk} \vert \rangle \neq 0$. In fact as we will show later frames of minimum Hubble variation are in large part determined by the bulk flows.

To sum up, in the standard framework, the cosmic rest frame and the frame of minimum Hubble flow variation need \emph{not} coincide for a particular observer. What we will now determine is whether the measured value of \vmin\ at our position \citep{Wilthvar,Wilthvar2} is consistent with the usual expectation. 

\subsection{Fitting the linear Hubble law}

Let us consider measuring the Hubble parameter in spherical shells about a particular point (here we follow the notation of \cite{Wilthvar}). The observer fits a linear Hubble law to data in each shell. We consider low redshifts where the linear Hubble law is a good approximation and there is no dependence on the parameters of the $\Lambda$CDM cosmological model (other than $H_0$).

The standard linear regression in each spherical shell $s$ is performed by minimising $\chi^2_\text{s}=\sum_i^{N_\text{s}}\left[ \sigma_i^{-1}(r_i - c z_i / H) \right]^2$ with respect to $H$, where $r_i$ is the distance to a particular object, $\sigma_i$ is the error on the distance measurement, $N_\text{s}$ is the number of objects\footnote{e.g. galaxies in a galaxy survey or halos in a N-body simulation} in the spherical shell, and $z_i$ is the redshift of the object. This gives the Hubble parameter for that shell and the associated error:

\begin{equation}
H_\text{s} = \left( \sum_{i=1}^{N_\text{s}} \frac{\left( c z_i \right)^2}{\sigma_i^2} \right) \left( \sum_{i=1}^{N_\text{s}} \frac{c z_i r_i}{\sigma_i^2} \right)^{-1},\quad \sigma_\text{s}^2 = \left( \sum_{i=1}^{N_\text{s}} \frac{\left( c z_i \right)^2}{\sigma_i^2} \right)^{3} \left( \sum_{i=1}^{N_\text{s}} \frac{c z_i r_i}{\sigma_i^2} \right)^{-4} .
\label{eq:Hs}
\end{equation}
In a N-body simulation there are no distance uncertainties, so $\sigma_i$ in the equation above can be set to unity for simplicity and without loss of generality. This yields the ``pure" $\Lambda$CDM result. Survey specific predictions can be obtained by producing mock catalogues which include the distance (and other) uncertainties.

We study the contributions to the Hubble parameter and, later, the effects of changing reference frames, by expanding the relevant quantities up to the dipole term. The redshifts of objects as measured in some initial frame in a thin spherical shell at distance $r$ are:

\begin{equation}
z_i=z(r)(1+\mathbf{d \cdot \hat{r}}_i + ...),
\label{eq:red}
\end{equation}
where $z(r)$ is the background redshift at the position of an object and $\mathbf{\hat{r}}_i$ is a unit vector in the direction of that object. The dipole part for the continuous distribution of objects can be estimated as
\begin{equation}
\mathbf{d} = 3 \langle z(\theta, \phi) \mathbf{\hat{r}}(\theta, \phi) \rangle / \langle z(\theta, \phi) \rangle,
\end{equation}
where $\theta$ and $\phi$ are spherical polar angles and $\langle \bullet \rangle$ corresponds to the spherical average. We use the fact that $\langle  \hat{r}^2_x\rangle=\langle  \hat{r}^2_y\rangle=\langle  \hat{r}^2_z\rangle=1/3$.
The formula above can be rewritten then for isotropically distributed discrete tracers as
\begin{equation}
\mathbf{d} \approx 3 \left( \sum_{i=1}^N z_i  \mathbf{\hat{r}}_i \right) / \left( \sum_{i=1}^N  z_i \right).
\label{eq:reddip}
\end{equation}
The redshifts can be written in terms of the peculiar velocities $\mathbf{v}_\mathrm{p}$ as (setting $c=1$):

\begin{equation}
z_i=z(r)+ \mathbf{v}_\mathrm{p} \cdot \mathbf{\hat{r}}_i.
\end{equation}
All velocities considered in this paper are non-relativistic hence we neglect relativistic effects which are of $\mathcal{O}(v/c)$ i.e. $\sim0.1\%$ for typical peculiar velocities of several hundred km~s$^{-1}$. This is much smaller than errors in distance measurements, so are irrelevant in practice when working with real data.

By applying Eq.(\ref{eq:reddip}) to the above expression we find that the dipole term is approximately related to the bulk velocity of the spherical shell: 
\begin{equation}
\mathbf{d}(r) \approx \mathbf{v}_\text{bulk}(r)/z(r) .
\label{eq:dipbulk}
\end{equation}
This expression is exact when all the objects have the same peculiar velocity.\footnote{The simple estimator (\ref{eq:reddip}) receives contributions from higher moments, so is \emph{not} reliable for estimation of bulk velocity. However, it is precisely this sum from Eq.(\ref{eq:reddip}) that is needed later on, e.g. in Eq.(\ref{eq:Hdiff}).}
The bulk velocity is defined here as the volume average of the peculiar velocities:
\begin{equation}
\mathbf{v}_\text{bulk}=\frac{1}{V}\int_{V} \mathbf{v}_\mathrm{p}\, \mathrm{d}V \approx \frac{1}{N} \sum_{i=1}^{N} \mathbf{v}_{p,i},
\end{equation}
where the last approximation is used to estimate the bulk flow of $N$ discrete tracers in volume $V$ (e.g. halos in a N-Body simulation).\footnote{This is the approach of Ref.\citep{LiBulk} which also showed that the bulk flow estimate does not depend on the mass of the halos used, so we can simply employ the number averages.}

The spherically averaged Hubble parameter obtained by a linear fit (\ref{eq:Hs}) is:
\begin{equation}
H_\text{s} = \frac{\langle (z(r)(1+\mathbf{d \cdot \hat{r}}_i + ...))^2 \rangle }{\langle (z(r)(1+\mathbf{d \cdot \hat{r}}_i + ...))r \rangle } = \frac{z(r)}{r}\left( 1 + \frac{1}{3}\vert \mathbf{d}\vert ^2 + ...\right) \approx \frac{z(r)}{r}\left( 1 + \frac{1}{3}\frac{\vert \mathbf{v}_\text{bulk}(r)\vert ^2}{z(r)^2}\right) ,
\label{eq:Hsd}
\end{equation}
where $\langle \bullet \rangle$ corresponds to the spherical average (e.g. the spherical average $\langle (\mathbf{d \cdot \hat{r}}_i)^2\rangle$ is $\vert \mathbf{d}\vert ^2 /3$). The linear terms in $H_\text{s}$ above on average cancel for an isotropic distribution of objects and we are left only with the quadratic contributions. The cancellation of linear terms works even when the sky coverage is incomplete provided the missing patches are symmetrically distributed on the opposite sides of the sky as is the case for e.g. the Zone of Avoidance.

Note that the Hubble parameter obtained by performing a linear fit of the Hubble diagram is biased towards higher values provided that the redshifts of objects have a non-zero dipole term or equivalently if the bulk flows are non-zero. The typical magnitude of the bulk flows is of order a few hundred km/s \citep{feldmanbulk,watkinsbulk,kashlinskybulk,kashlinskybulk2,watkinsbulk2,colinbulk,nusserbulk,mabulk,turnbullbulk} hence below $\sim70$\mpc\ where $H_0 r \approx 7000$\,km/s, the bias can be up to a few percent but becomes negligible for larger distances. The precise value of this bias can be calculated in linear theory or, more accurately, estimated from N-Body simulations since the distances involved are already in a mildly non-linear regime.

\subsection{Boosted frames and systematic offset of $H_\text{s}$}
\label{subsec:boosts}
Now we boost an observer by $\mathbf{v}$ with respect to the initial frame. The redshifts change to:

\begin{equation}
z'_i=z_i-\mathbf{v \cdot \hat{r}}_i=z(r)(1+\mathbf{d' \cdot \hat{r}}_i + ...),
\end{equation}
where  $\mathbf{d'}=\mathbf{d}-\mathbf{v}/z(r)$. Using Eq.(\ref{eq:Hs}) and following Eq.(\ref{eq:Hsd}) we find the difference between the spherically averaged Hubble parameters in boosted and initial frames to be:

\begin{equation}
H_\text{s}' - H_\text{s} = \frac{z(r)}{r}\left( 1 + \frac{1}{3}\vert \mathbf{d'}\vert ^2 + ...\right) - \frac{z(r)}{r}\left( 1 + \frac{1}{3}\vert \mathbf{d}\vert ^2 + ...\right)  = \frac{\frac{1}{3}\vert \mathbf{v} \vert^2 - \frac{1}{3} 2 \mathbf{v} \cdot \mathbf{d}(r)z(r) + ...}{\langle z(r)r \rangle}.
\label{eq:Hdiff}
\end{equation}
%
%\begin{equation}
%H_\text{s}' - H_\text{s} = \frac{\langle (z(r)(1+\mathbf{d' \cdot \hat{r}}_i + ...))^2 \rangle }{\langle (z(r)(1+\mathbf{d' \cdot \hat{r}}_i + ...))r \rangle } - \frac{\langle (z(r)(1+\mathbf{d \cdot \hat{r}}_i + ...))^2 \rangle}{\langle (z(r)(1+\mathbf{d \cdot \hat{r}}_i + ...))r \rangle} = \frac{\frac{1}{3}\vert \mathbf{v} \vert^2 - \frac{1}{3} 2 \mathbf{v} \cdot \mathbf{d}(r)z(r) + ...}{\langle z(r)r \rangle}.
%\label{eq:Hdiff}
%\end{equation}
%
%\begin{equation}
%H_\text{s}' - H_\text{s} \approx \frac{\frac{1}{3}\vert \mathbf{v} \vert^2 - \frac{1}{3} 2 \mathbf{v} \cdot \mathbf{v}_\text{bulk}(r)}{H_0 \langle r^2 \rangle},
%\label{eq:Hdiff2}
%\end{equation}
Furthermore, we can replace $z(r)$ with $H_0 r$. The error in this substitution is of order $\delta H$ so it affects the equation above (already of order $\delta H$) at the next order which we neglect. We can also approximate $\mathbf{d}(r)$ with $\mathbf{v}_\text{bulk} (r) / z(r)$ as in Eq.(\ref{eq:dipbulk}). This leads to
\begin{equation}
H_\text{s}' - H_\text{s} \approx \frac{\frac{1}{3}\vert \mathbf{v} \vert^2 - \frac{1}{3} 2 \mathbf{v} \cdot \mathbf{v}_\text{bulk}(r)}{H_0\langle r^2 \rangle}.
\label{eq:Hsbulk}
\end{equation}

We stress that any initial frame, including the frame in which the Hubble flow is most uniform, can have and typically does have dipole structure in the velocity field, as we show later. Allowing for this dipole structure originating in bulk flows leads to the second term in the numerator in Eq.(\ref{eq:Hdiff}) above. This term spoils the pure $1/\langle r^2 \rangle$ dependence of the differences in Hubble parameters in spherical shells between boosted and initial frames. Our formula (\ref{eq:Hdiff}) for systematic offset of $H_\text{s}$ between reference frames is thus \emph{different} from corresponding equations in Refs.\citep{Wilthvar,Wilthvar2} where the second dipole term was neglected under the assumption that the initial frame has no dipole structure.\footnote{In addition, there is a small numerical error in the denominator of Eq.(9) of Ref.\citep{Wilthvar} where the factor 2 should be 3, and similarly in Refs.\citep{Wilthvar2,Bolejkodiff}.}  Therefore, the refinements of Eq.(\ref{eq:Hdiff}) and the correct spherical averages of quantities need to be taken into account when studying the systematic offsets in $H_\text{s}$ due to Lorentz boosts and assessing the size of the non-kinematical effects studied in Refs.\citep{Wilthvar,Wilthvar2}.

Boosts can either increase or decrease the Hubble parameter in each spherical shell. Whereas $H_\text{s}'$ can be increased without bound with large enough boost $\mathbf{v}$, there is a limit to how much it can be decreased. Let us write the boost in terms of the components parallel and perpendicular to the bulk velocity of the objects in the spherical shell, $\mathbf{v}=\alpha \mathbf{v}_\text{bulk}(r) + \mathbf{v}_{\perp bulk}(r)$. Minimising Eq.(\ref{eq:Hsbulk}) results in
\begin{equation}
\text{Min} \left( H_\text{s}' - H_\text{s} \right) \approx \text{Min} \left( \frac{\frac{1}{3}(\alpha^2 \vert \mathbf{v}_\text{bulk} \vert^2 + \vert \mathbf{v}_{\perp \text{bulk}} \vert^2)  - \frac{1}{3} 2 \alpha \vert \mathbf{v}_\text{bulk} \vert^2}{H_0 \langle r^2 \rangle} \right) = \frac{-\frac{1}{3}\vert \mathbf{v}_\text{bulk}(r) \vert^2}{H_0 \langle r^2 \rangle}.
\label{eq:minhs}
\end{equation}
This corresponds to the boost to the frame where the dipole vanishes and the Hubble parameter $H_\text{s}$ only receives contribution from the pure monopole $z(r)/r$ (see Eq.\ref{eq:Hsd}). Note that the dipoles are in general \emph{different} from shell to shell and a single boost cannot make the dipoles vanish in all the shells simultaneously.

%Note that if we start with a pure monopole flow (i.e. $\mathbf{d}=\mathbf{v}_\text{bulk}=0$), a dipole introducing boost can only increase the spherically averaged Hubble parameter. 

It follows that there are two separate cases based on whether the monopole part of the flow before the boost is above or below the background value $H_0$. When it is below, a boost can bring $H_\text{s}$ arbitrarily close to $H_0$. If it is above, $H_\text{s}$ can only be brought down to its monopole value which is the lowest any boost can achieve.

%The Hubble parameter itself, again in a thin spherical shell at distance $r$, can be expanded up to the dipole term:
%\begin{equation}
%H(\mathbf{r})=H(r)(1+\mathbf{d}_H \cdot \mathbf{\hat{r}} + ...)
%\end{equation} 
%which upon using $H(\mathbf{r})=cz(\mathbf{r})/r$ and Eq.(\ref{eq:red}) yields $\mathbf{d}_H=\mathbf{d}$.

\subsection{Finding the frame of minimum Hubble variation}
 We now define following Ref.\citep{Wilthvar} the measure of closeness of Hubble flow to the expected background flow and the method of finding the frame of the most uniform Hubble flow.

The deviation of the Hubble parameter in a spherical shell from the background value is $\delta H_\text{s} = (H_\text{s} - H_0)$. The case of uniform flow is $\delta H=0$, which is also the ensemble expectation of linear perturbation theory.
The frame with minimum Hubble flow variation for a particular observer is found by solving for a boost velocity $\mathbf{v}=\mathbf{v}_\text{min}$ that minimises the sum of the mean square differences of $\delta H$, weighted by their errors:
\begin{equation}
\chi^2 (\mathbf{v})= \sum_{j=s}^{N_\text{shells}} \frac{\delta H_j^2}{\sigma_{H_j}^2}.
\label{eq:chisq}
\end{equation}
The boost velocity $\mathbf{v}_\text{min}$ will depend somewhat on the details of binning the distances into concentric shells. The expression above is simplified compared to Refs.\citep{Wilthvar,Wilthvar2} --- in a N-Body simulation the value of $H_0$ is known exactly and so its errors are not included above.

We briefly mention another method to estimate the Hubble parameter that is suitable for, and frequently used in, N-body simulations where there are no errors on distances $r_i$. It is using an estimator that is linear in redshift:
\begin{equation}
H_{s, \mathrm{lin}} = \frac{1}{N_\text{s}}\sum_{i=1}^{N_\text{s}} \frac{c z_i}{r_i}, \quad \sigma_\text{s}^2 = \frac{1}{N_\text{s}}\sum_{i=1}^{N_\text{s}} \left( \frac{c z_i}{r_i} \right)^2 - H_{s, \mathrm{lin}}^2
\label{eq:Hlin}
\end{equation}
The value of $H_{s,\mathrm{lin}}$ is unaffected by boosts because the linear contribution will cancel on average in Eq.(\ref{eq:Hlin}) above. Hence the velocities of frames of minimum Hubble variation have a trivial uniform distribution when using the estimator above. In this paper we use the observationally relevant estimator in Eq.(\ref{eq:Hs}).

\subsection{Linear perturbation theory}
\label{sec:linear}

Linear perturbation theory provides some insight into the relevant factors that determine the frames of minimum Hubble variation within the framework of standard cosmology.

The peculiar velocity field can be written as:

\begin{equation}
\mathbf{v}_\mathrm{p}(\mathbf{r},t)=\frac{H(t)  f}{4 \pi} \int \frac{\delta(\mathbf{r}')(\mathbf{r}-\mathbf{r}')}{|\mathbf{r}-\mathbf{r}'|^3}\, \mathrm{d}\mathbf{r}'^3 ,
\label{eq:linvel}
\end{equation}
where $H(t)$, $f$, $\delta(\mathbf{r})$ are the background Hubble parameter, the scale factor, the growth factor, and the overdensity respectively. Note that the dominant contribution in Eq.(\ref{eq:linvel}) comes from large scales.  
Taking the divergence of the peculiar velocity field gives:
\begin{equation}
\nabla \cdot \mathbf{v}_\mathrm{p}(\mathbf{r},t) = -H(t)  f \delta(\mathbf{r}).
\end{equation}
The local Hubble parameter is proportional to the divergence of the velocity field:
\begin{equation}
H_\text{loc}(\mathbf{r},t)=H(t)+\frac{1}{3} \nabla \cdot \mathbf{v}_\mathrm{p}(\mathbf{r},t) =H(t)-\frac{1}{3}H(t)  f \delta(\mathbf{r}).
\label{eq:Hdens}
\end{equation}

The bulk flow velocity \vb\ is obtained by taking a volume average of the peculiar velocity field in Eq.(\ref{eq:linvel}).  Note that for spherical volumes the density monopole does not affect the bulk velocity in linear theory, that is, the dominant contribution comes from the density dipole term. 
However, the value of the Hubble parameter is positively correlated with the density monopole (i.e. low/high values correspond to overdensities/underdensities), as has been noted in the context of N-Body simulations \citep{HubbleNBody}.

We see from Eq.(\ref{eq:Hdiff}) that the contribution of a boost (consequently \vmin) to the linear fit estimate of the Hubble parameter (\ref{eq:Hs}) depends on the bulk flow, \vb, as well as on the density monopole (via Eq.\ref{eq:Hdens}). This leads to a non-trivial  dependence of \vmin\ on the density of the local environment as well as the bulk flow as measured in the CMB frame.

\subsection{Finite Infinity}

The notion of finite infinity (FI) was originally introduced in discussion of the `fitting problem' where it was defined as a time-like surface within which the dynamics can be treated as isolated from the rest of the universe \citep{EllisFI}. Here, we adopt a more specific definition in line with Refs.\citep{WiltFI, WiltFI2}. Finite infinity is associated with the smallest region within which the average expansion, $\langle \theta \rangle_\text{V}$, vanishes, while being positive outside. That is, the boundaries of finite infinity separate collapsing regions from the expanding ones. We further simplify the definition by estimating the average expansion, or equivalently the divergence of the velocity field, in a spherical volume using discrete tracers (i.e. DM halos):
\begin{equation}
\langle \theta \rangle_\text{V} = \langle \nabla \cdot \mathbf{v} \rangle_\text{V} \approx \frac{3}{N} \sum_{i=1}^N \left( \frac{\mathbf{v}_i \cdot \mathbf{\hat{r}}_i}{r_i} \right),
\end{equation}
where $\mathbf{v}=H_0 \mathbf{r} + \mathbf{v}_\mathrm{p}$. The region of finite infinity has an associated bulk(group) velocity, $\mathbf{v}_\text{fi}$, with respect to the CMB frame. Virialised structure and discreteness of the tracers provide the lower cut-off to the size and meaningfulness of the FI region.

\subsection{A simple picture}

Let us illustrate the relations among various velocities discussed so far with a simple set-up. Imagine a spherical volume of space that has a small collapsing central part while the outer part is expanding with the background (i.e. uniform Hubble flow). On top of this flow let us add peculiar velocities that are the same for each tracer, $\mathbf{v}_\text{p}$. The bulk flow for any spherical subvolume or shell is $\mathbf{v}_\text{bulk}=\mathbf{v}_\text{p}$. In the CMB frame this bulk flow will bias the linear fit estimate of the Hubble parameter in spherical shells (see Eqs.(\ref{eq:Hs}) and (\ref{eq:Hsd})). Boosting into a frame with $\mathbf{v}_\text{min}=\mathbf{v}_\text{p}$ will remove the bulk flow and restore the background Hubble parameter in the outer regions. At the same time the bulk velocity of the collapsing finite infinity region,$\mathbf{v}_\text{fi}$, is the same as the bulk flow of the whole spherical region and the velocity of the frame of most uniform Hubble flow ($\mathbf{v}_\text{min}=\mathbf{v}_\text{bulk}=\mathbf{v}_\text{fi}$). In this simple situation we have a trivial peculiar velocity profile (i.e. constant and perfectly correlated). Since the actual velocity field is correlated, this simplistic set-up is still relevant hence we expect that the directions and magnitudes of $\mathbf{v}_\text{min},\mathbf{v}_\text{bulk},\mathbf{v}_\text{fi}$ are also correlated.

\section{Results and Discussion}
\label{sec:results}

When working with simulation data we use radial bins of width 10\mpc\ with the lower limit 10\mpc\ and the upper limit 100\mpc. Changing the bin size (to 20\mpc) and lower and upper limits (to 20\mpc\ and 120\mpc\ respectively) do not affect the results by more than 10\%. We use the simulation output at $z=0$ as we are only interested in distances below 100\mpc\ (corresponding to $z\approx0.025$) where the Hubble law is linear. %and there is no need to include quadratic corrections in $z^2$.

We select observers randomly in the simulation volume. This simple procedure does not place observers in halos where they may be expected to be in the actual universe \citep{TurnerNBody}. However our volume weighted approach enables direct comparison to perturbation theory and analytical calculations (e.g. Refs.\citep{MarraHvar,wanglin}). We define overdense and underdense regions  up to 100\mpc~as those that are consistently over or under the average density.

\subsection{Systematic offset in Hubble parameter in different reference frames}

The larger monopole variation of the Hubble law in the CMB frame compared to the LG frame as was noted in Refs.\citep{Wilthvar,Wilthvar2} is due to the non-linear dependence of $H_\text{s}$ on the redshift (see Eq.\ref{eq:Hs}). Boost to a different frame results in a systematic offset in the Hubble parameter $H_\text{s}$. We have improved the earlier calculation \citep{Wilthvar,Wilthvar2} by recognising that the frame of minimum Hubble flow variation has a dipolar structure. This yields the expression (\ref{eq:Hdiff}) which shows that the systematic offsets in $H_\text{s}$ do \emph{not} have the pure $1/r^2$ dependence assumed earlier, but additionally have a dependence on the dipole structure of the velocity field.
 
\begin{figure}
\centering
 \includegraphics[width=0.5\columnwidth]{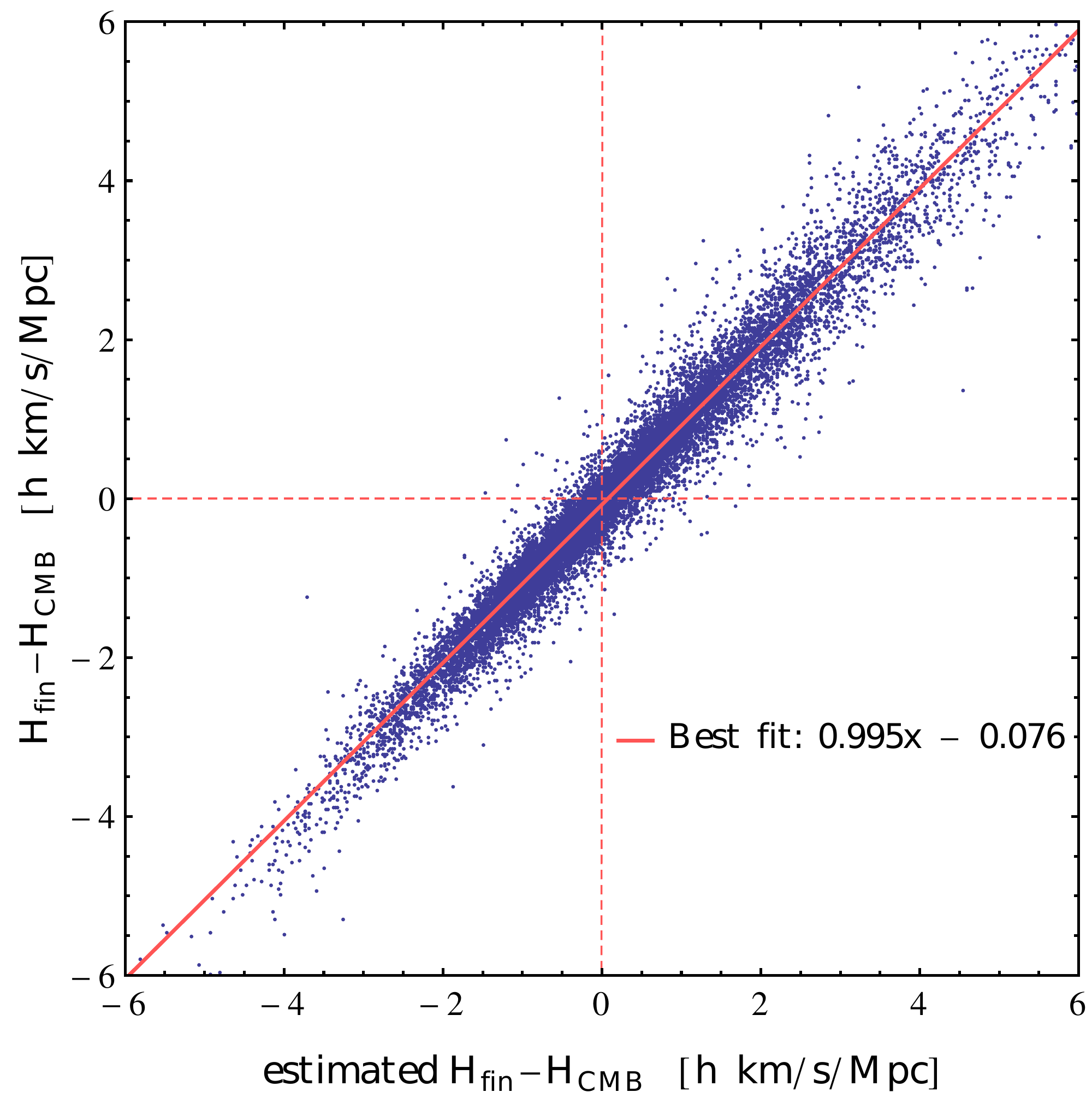}
 \caption{Actual versus estimated $H_\text{fin}-H_\text{CMB}$ for the radial bin at 40-50\mpc.} %The best fit line is plotted.}
   \label{fig:predHs}
\end{figure}

Using a N-Body simulation we extracted the difference between the Hubble parameters in the minimum Hubble variation frame and the CMB frame, $\delta H_\text{s} = H_\text{s,fin}-H_\text{s,CMB}$, by looking for the appropriate boost \vmin\ using Eq.(\ref{eq:chisq}). This difference $\delta H_\text{s}$ was compared to the estimate (\ref{eq:Hdiff}) where the dipoles in each spherical shell were obtained using Eq.(\ref{eq:reddip}). We plot the actual against the estimated $\delta H_\text{s}$ in Fig.\ref{fig:predHs} which shows that the points are well correlated with the best-fit slope very close to unity. Additionally, the scatter of the points provides an estimate of the error in Eq.(\ref{eq:Hdiff}) due to neglecting the terms higher than dipole.

We now test the boost formula (\ref{eq:Hdiff}) on real data in order to understand the discrepancy between the measured value of $H_\text{CMB}-H_\text{LG}$ and the $1/r^2$ expectation from Refs.\citep{Wilthvar,Wilthvar2}.  We use both the \textit{Cosmicflows-2.1} catalogue \citep{cosmicflows} and the \textsc{composite} catalogue \citep{feldmanbulk,watkinsbulk} to obtain the Hubble parameters in spherical shells in the CMB and LG frames, %closely following Ref.\citep{Wilthvar2}.
The dipoles needed for our estimation of $H_\text{CMB}-H_\text{LG}$ in Eq.(\ref{eq:Hdiff}) are obtained via Eq.(\ref{eq:reddip}). The errors in the estimation are a combination of the distance errors in the catalogues and the error in the formula itself extracted from the tests on our N-Body simulation. Note that \textit{Cosmicflows-2.1} has a Malmquist bias which largely cancels in $H_\text{CMB}-H_{LG}$, however, some small differences in the measured values between the two catalogues can be attributed to this bias \citep{Wilthvar2}.

\begin{figure}
\centering
\begin{subfigure}[b]{0.5\columnwidth}
 \includegraphics[width=\columnwidth]{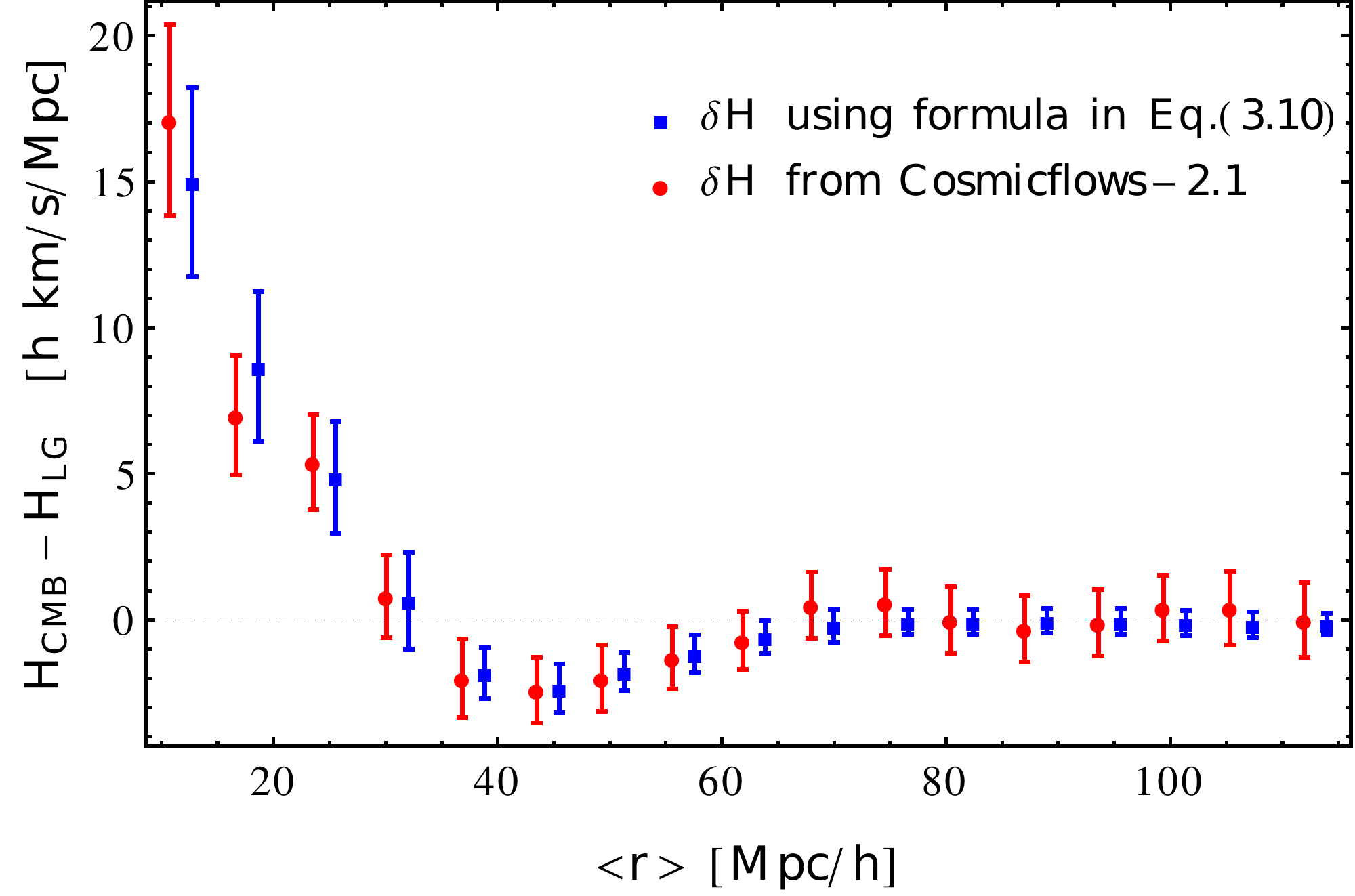}   
\end{subfigure}%
~~
\begin{subfigure}[b]{0.5\columnwidth}
 \includegraphics[width=\columnwidth]{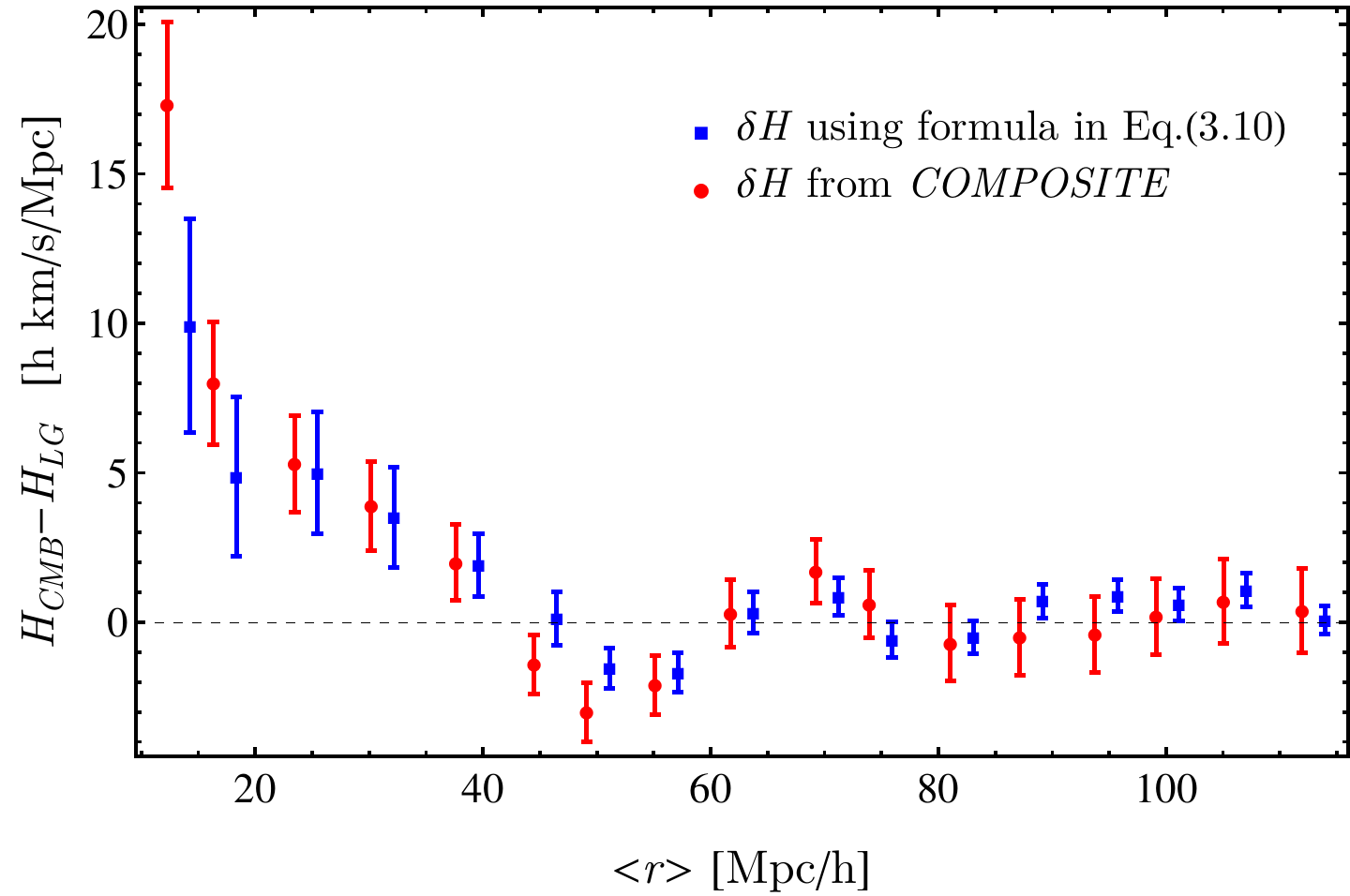}
\end{subfigure}%

\caption{Actual vs estimated $H_\text{CMB}-H_\text{LG}$ for the \textit{Cosmicflows-2.1} and \textsc{composite} catalogues. The predicted values have the radial coordinate shifted by 2\mpc\ for clarity. The average radial coordinate of a bin, $\langle r \rangle$,  is the error weighted average of positions of objects in that bin.}
\label{fig:deltaH}
\end{figure}

In Fig.\ref{fig:deltaH} we plot $H_\text{CMB}-H_{LG}$ in radial bins chosen as in Refs.\citep{Wilthvar,Wilthvar2}. Note that the differences in Hubble parameters in the CMB and LG frames are completely consistent with the boost formula (\ref{eq:Hdiff}). The previously highlighted discrepancy was mainly due to the assumption that the frames close to the minimum Hubble flow variation frame have negligible dipole structure of the velocity field. However, as discussed in \S~\ref{subsec:boosts}, the initial dipoles in spherical shells are typically different (albeit correlated), and a boost to the frame of minimum Hubble flow variation therefore never reduces all the dipoles to zero. In fact, if the initial Hubble parameters in spherical shells are below the background value, then the boost to the minimum Hubble flow variation frame will \emph{increase} the dipoles. Therefore, the dipole structure must necessarily be considered when deriving the expectation for the systematic offsets in $H_\text{s}$ due to change of reference frame.

\subsection{Variation of $H$}

\begin{figure}
\centering
 \includegraphics[width=0.5\columnwidth]{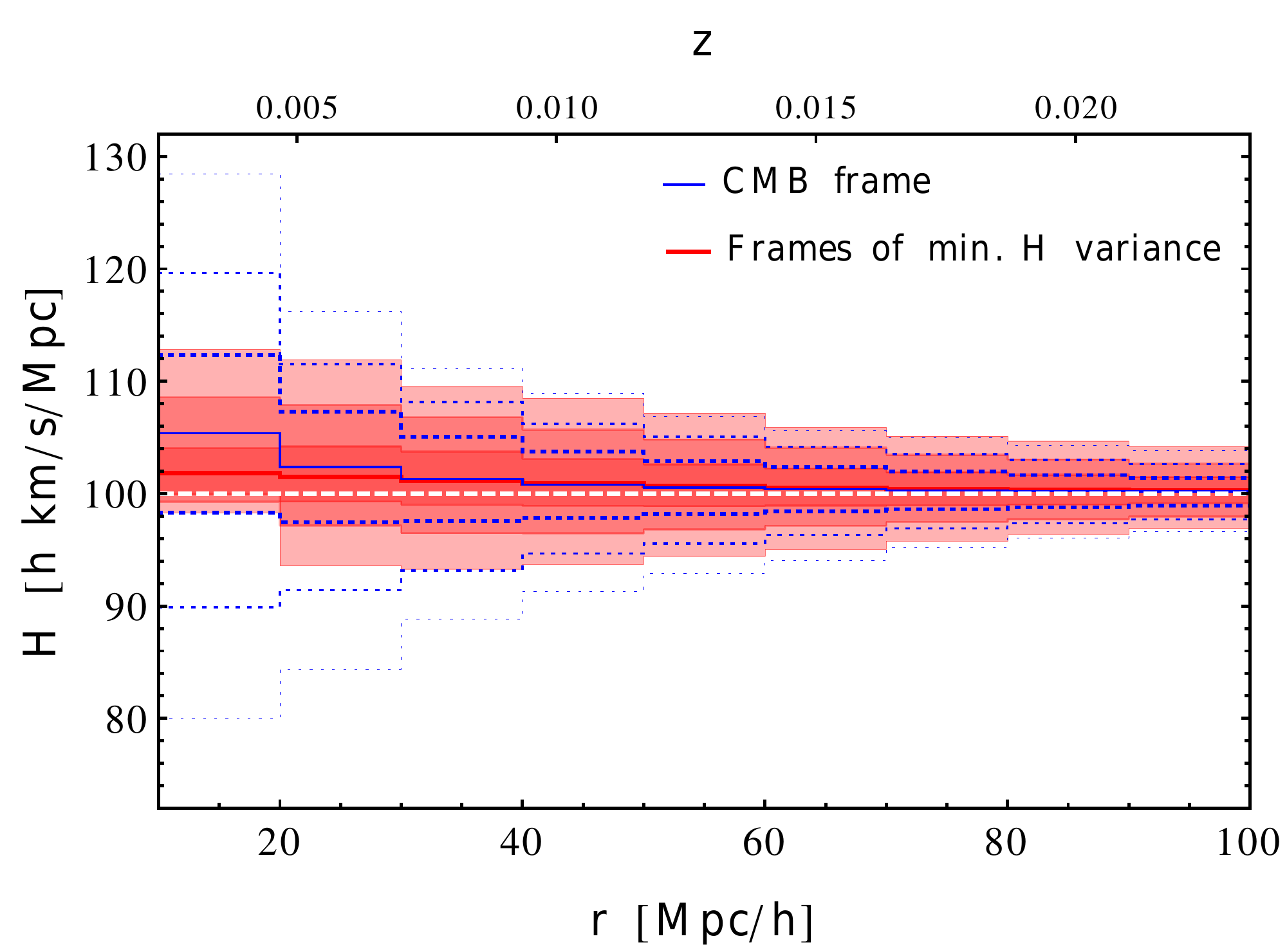}
 \caption{Hubble parameter as measured by randomly placed observers in the CMB frame (blue lines) and as measured in frames of most uniform Hubble flow (shaded red). The contours are 68.3\%, 95.4\% and 99.7\% confidence intervals while solid lines are the mean values. Note the bias towards higher values of $H$ compared to $H_0$.}
   \label{fig:Hvar}
\end{figure}

In this section we study the effect of boosting into frames with minimal Hubble variation. The expectation is that such boosts bring the radial Hubble flow profile closer to the background expectation of $H_0 \equiv 100h$~km/s/Mpc. Indeed we see in Fig.\ref{fig:Hvar} that the spread of measured Hubble parameter is much smaller in every radial bin and closer to $H_0$, once every observer is boosted by their respective \vmin. However, above $\sim60$\mpc\ there is not much difference. The effect of changing frames is suppressed by $1/\langle r^2 \rangle$ (see Eq.\ref{eq:Hdiff}), so larger distances are less important in determining \vmin\ and less affected by the boosts. Therefore, most of the influence on the \vmin\ velocity distribution comes from the regions in the mildly non-linear to non-linear regime, which justifies the use of a N-Body simulation in our work.

At late times most of the volume of the universe is in voids. If an observer is placed in a random location, it is more likely to be located in an underdensity than an overdensity. This leads to the bias of the Hubble parameter towards higher values compared to $H_0$ as can be seen in Fig.\ref{fig:Hvar} where the solid line represents the average value of $H$ for the ensemble of observers. This has also been noted in Ref.\citep{HubbleNBody}. Additionally, at scales below about 50\mpc\, there is an additional bias towards higher $H_\text{s}$ due to the dipole component of the velocity field as discussed in \S~\ref{sec:frames} and shown in Eq.(\ref{eq:Hsd}).

Boosting to frames of most uniform Hubble flow will bring the Hubble parameter closer to $H_0$, however, the monopole part cannot be reduced (see Eqs.\ref{eq:Hsd},\ref{eq:Hdiff},\ref{eq:minhs}) and therefore some residual bias towards higher values of $H$ is left also in those frames (again see Fig.\ref{fig:Hvar}). 

\begin{figure}
\centering
 \includegraphics[width=0.5\columnwidth]{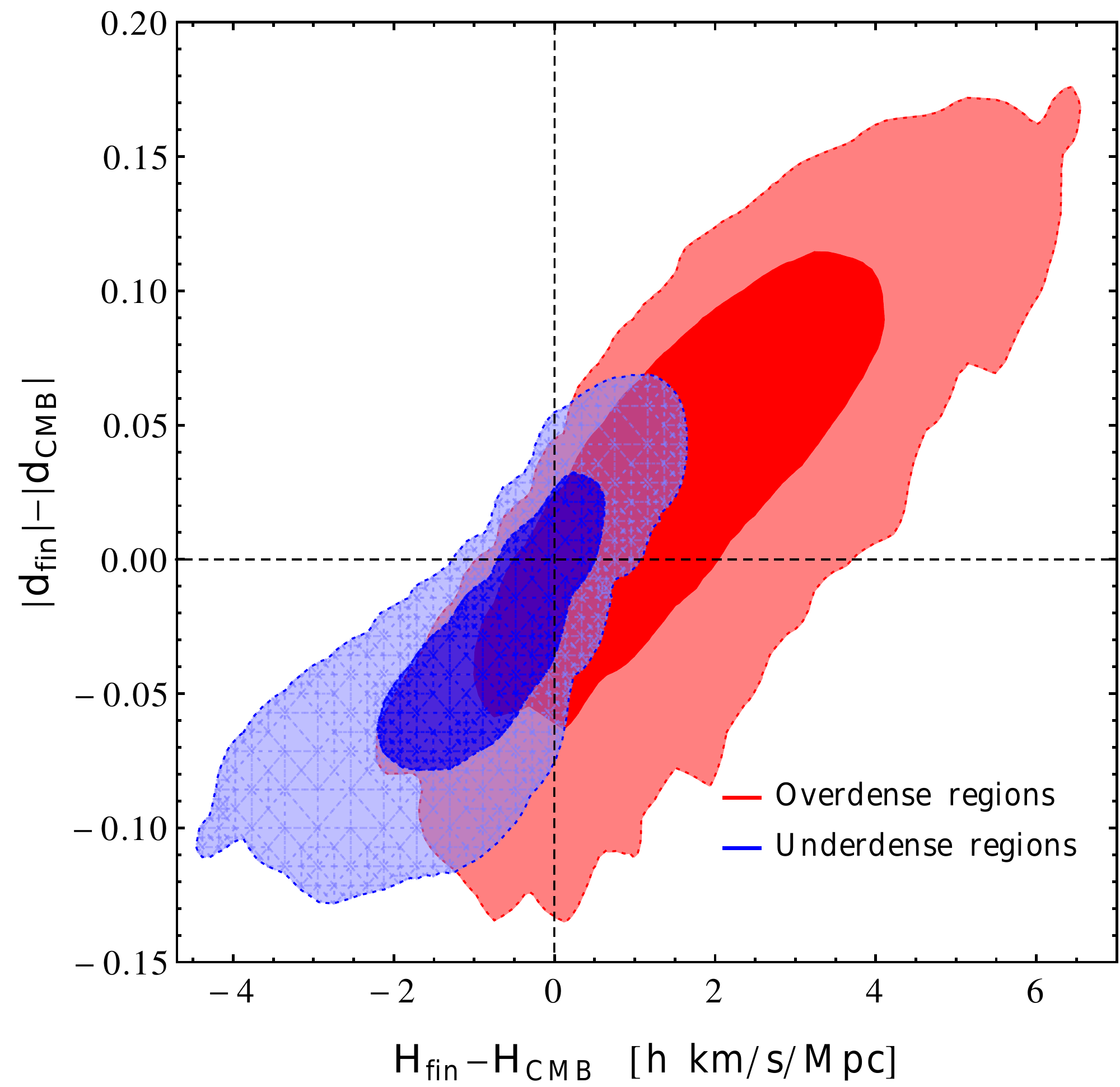}
 \caption{Difference between the dipole in the Hubble flow in the frame of minimum Hubble variation and in the CMB frame, plotted against the difference of the spherically averaged Hubble parameter. Observers in both overdense and underdense locations are chosen in the $40-50$\,\mpc\ radial bin. The contours correspond to 68.3\% and 95.4\% confidence intervals.}
   \label{fig:dvsH}
\end{figure}

For the cases where the value of $H$ is smaller than the background, which are on average associated with the overdense regions, boosts to the frame of minimum Hubble variation increase the dipole term in order to bring $H_\text{s}$ closer to $H_0$. Differently, in the underdense regions, boosts to the frame of minimum Hubble variation decrease the dipole term in order to bring $H_\text{s}$ closer to $H_0$. This preference can be seen in Fig.\ref{fig:dvsH} where we plot the difference between the dipoles against the difference in the Hubble parameters before and after boosting by \vmin\ for the over- and underdense positions.

\subsection{The probability distribution of \vmin}

\begin{figure}
\centering
 \includegraphics[width=0.5\columnwidth]{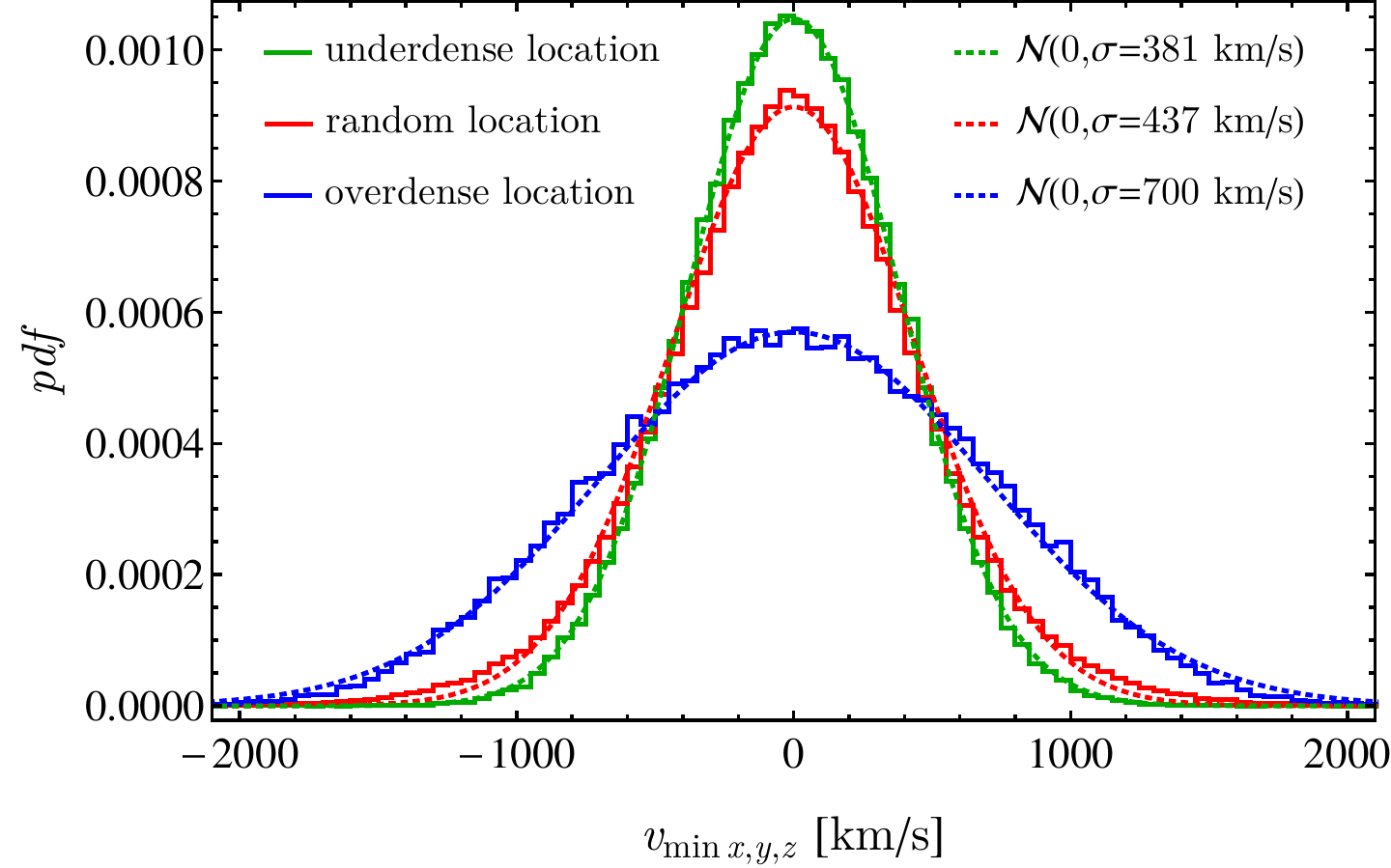}
 \caption{Probability distribution function for the components of the velocity  $\mathbf{V}_\text{min}$ for frames in which the Hubble flow looks most uniform. Observer locations are chosen based on the average density within 100 \mpc. The dashed lines are Gaussian fits.}
   \label{fig:Vmin}
\end{figure}

\begin{figure}
\centering
 \includegraphics[width=0.5\columnwidth]{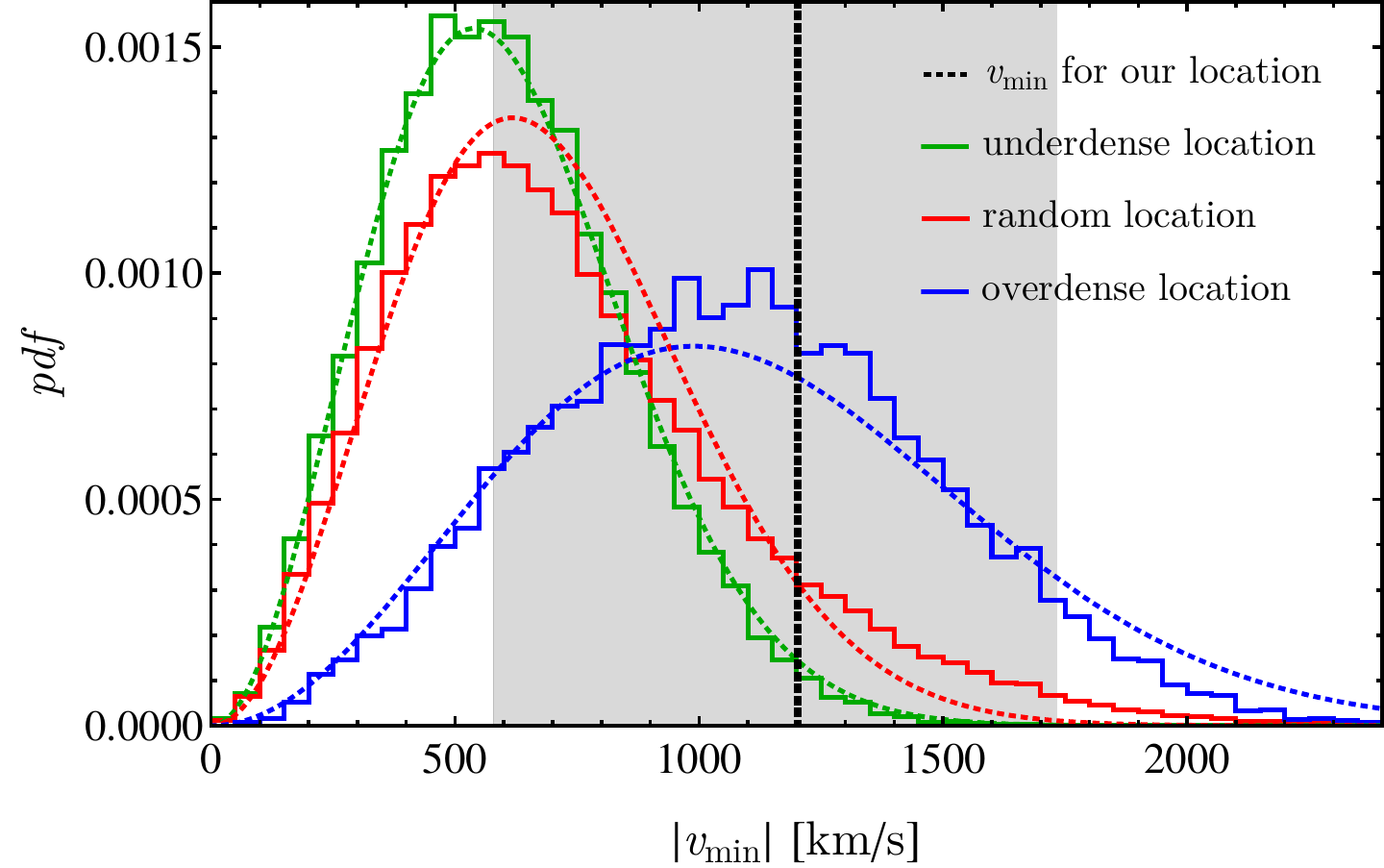}
 \caption{Probability distribution function for the norms of the velocity of frames of minimum Hubble variation. Observer locations are chosen as in the previous figure.}
   \label{fig:Vminnorm}
\end{figure}

Now we study the distribution of velocities \vmin\ of frames where the spherically averaged Hubble flow looks most uniform. We separate observers based on their local environment: random in volume, in underdensities, and in overdensities. We find \vmin\ by an algorithm that scans randomly for the velocity vector that minimises the variation (see Eq.\ref{eq:chisq}) with resolution of about 150\,km/s. The algorithm is then nested to achieve an average precision of 10\,km/s. Fig.\ref{fig:Vmin} shows the distribution of the individual components of \vmin --- the distributions are identical for each Cartesian component and centred on \vmin $=\mathbf{0}$ as required by the symmetries of the simulation. We perform a Gaussian fit to find the associated variance of \vmin and plot its norm in Fig.\ref{fig:Vminnorm}. This should be Maxwell-Boltzmann distributed if the underlying distribution for the components is truly Gaussian but we note there are small departures consistent with the variances of the Gaussian fits. We also mark the norm and the error of the boost that makes the Hubble flow around our position most uniform ($1203^{+529}_{-625}$ \,km/s) as obtained in Ref.\citep{Wilthvar2}). We note that this value is not inconsistent with the $\Lambda$CDM expectation for a randomly located observer. An earlier estimate \citep{Wilthvar} suggested that \vmin\ for our location approximately corresponds to the velocity of the Local Group ($635 \pm 38$\, km/s as inferred from the CMB dipole).

The amplitude of \vmin\ depends on the local environment in terms of average density as can be seen from Fig.\ref{fig:Vminnorm}. The denser the region around an observer's location the larger the amplitude of the boost. There are two reasons for this. Firstly, the Hubble parameter has bigger variance in the overdense locations relative to the underdense locations, stemming from the bigger variance in the density field there.\footnote{The underdense regions are bounded by $-1\leq \delta$ where $\delta$ is the fractional density. There is no such bound for overdense locations.} The bigger variance in $H$ will result in a bigger amplitude of boosts required. Secondly, in the overdense locations the monopole of the Hubble flow is on average smaller than $H_0$. The values of $H$ smaller than $H_0$ can be brought arbitrarily close to $H_0$ by introducing an extra dipole which results in bigger boosts compared to the underdense case. For underdense locations the Hubble parameter $H$ is expected to be above $H_0$. There is a limit, derived in Eq.(\ref{eq:minhs}), on how much the Hubble parameter can be lowered by a boost --- it can only be brought to its `pure' monopole value (which is on average closer to but still above $H_0$). Therefore the amplitude of \vmin\ for underdense locations is expected to be smaller compared to overdense locations.

\subsection{Correlations between \vb, \vmin, and $\mathbf{v}_\text{fi}$}

\begin{figure}
\centering
 \includegraphics[width=0.5\columnwidth]{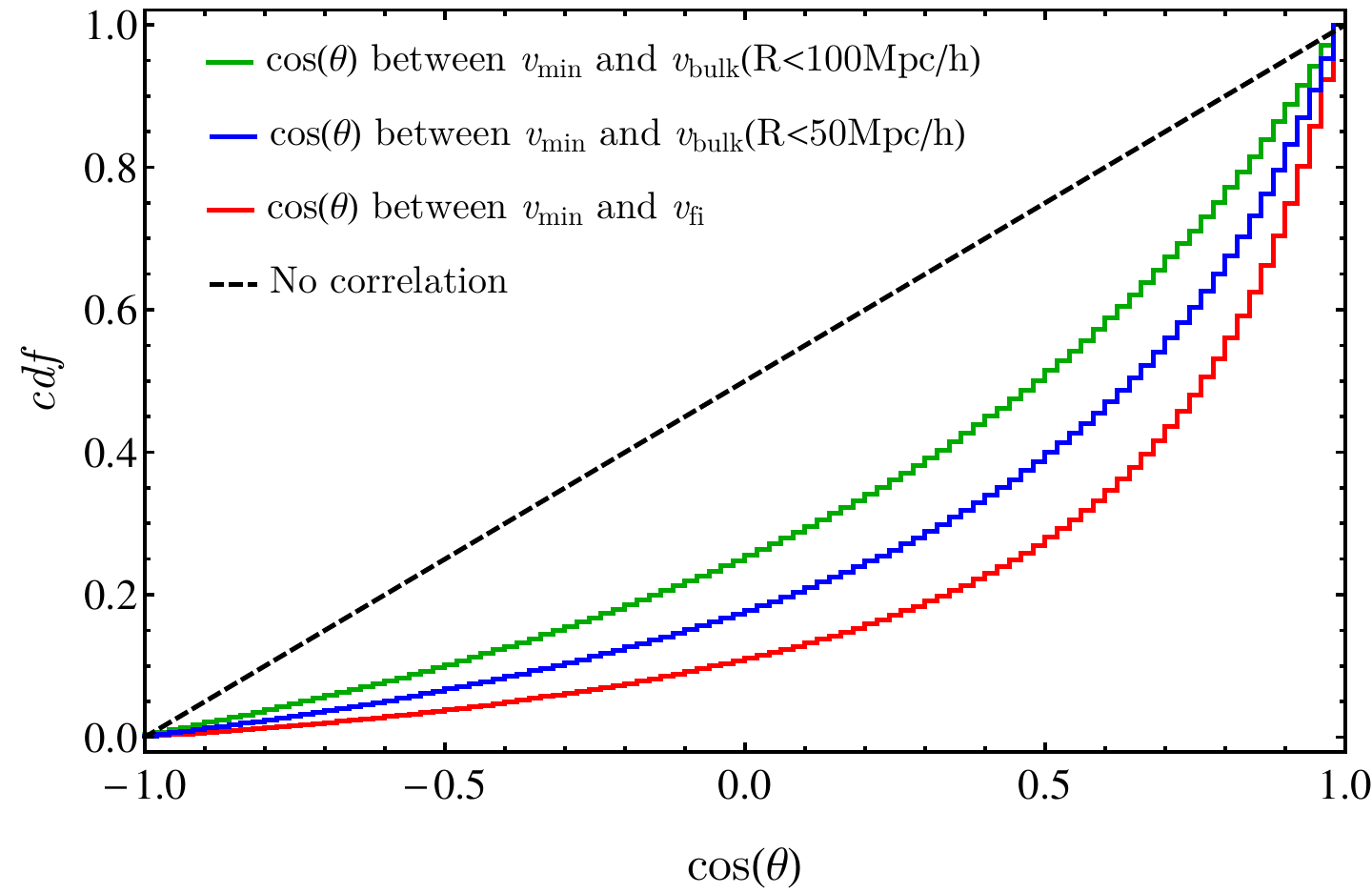}
 \caption{Cumulative distribution function for the cosine of the angle between $v_\text{min}$ and various bulk velocities.}
   \label{fig:cosv}
\end{figure}

In Fig.\ref{fig:cosv} we plot the cumulative distribution function (CDF) for the cosine of the angle between the boost \vmin\ and the bulk velocity of the finite infinity region $\mathbf{v}_\text{fi}$, as well as the bulk velocities within 50\mpc\ and 100\mpc\ radii. The observers are located at random in the simulation volume. In all cases the direction of \vmin\ is positively correlated with the bulk velocities. This is unsurprising as the average observer is located in an underdensity where boosts to the frames of minimum Hubble flow variation mainly reduce the existing dipole originating from the bulk velocity. The correlation is weaker for larger radii as the influence on \vmin\ is roughly suppressed by $1/\langle r^2 \rangle$ (see Eq.\ref{eq:Hdiff}) and is thus strongest for the region of `finite infinity' which is the smallest region considered (of typical size 15\mpc\ in our N-body simulation). This supports the expectation that the boost to the frame of minimum Hubble variation should roughly correspond to the frame of the bulk motion of the `finite infinity' region \citep{Wilthvar2}. 

\begin{figure}
\centering
 \includegraphics[width=0.5\columnwidth]{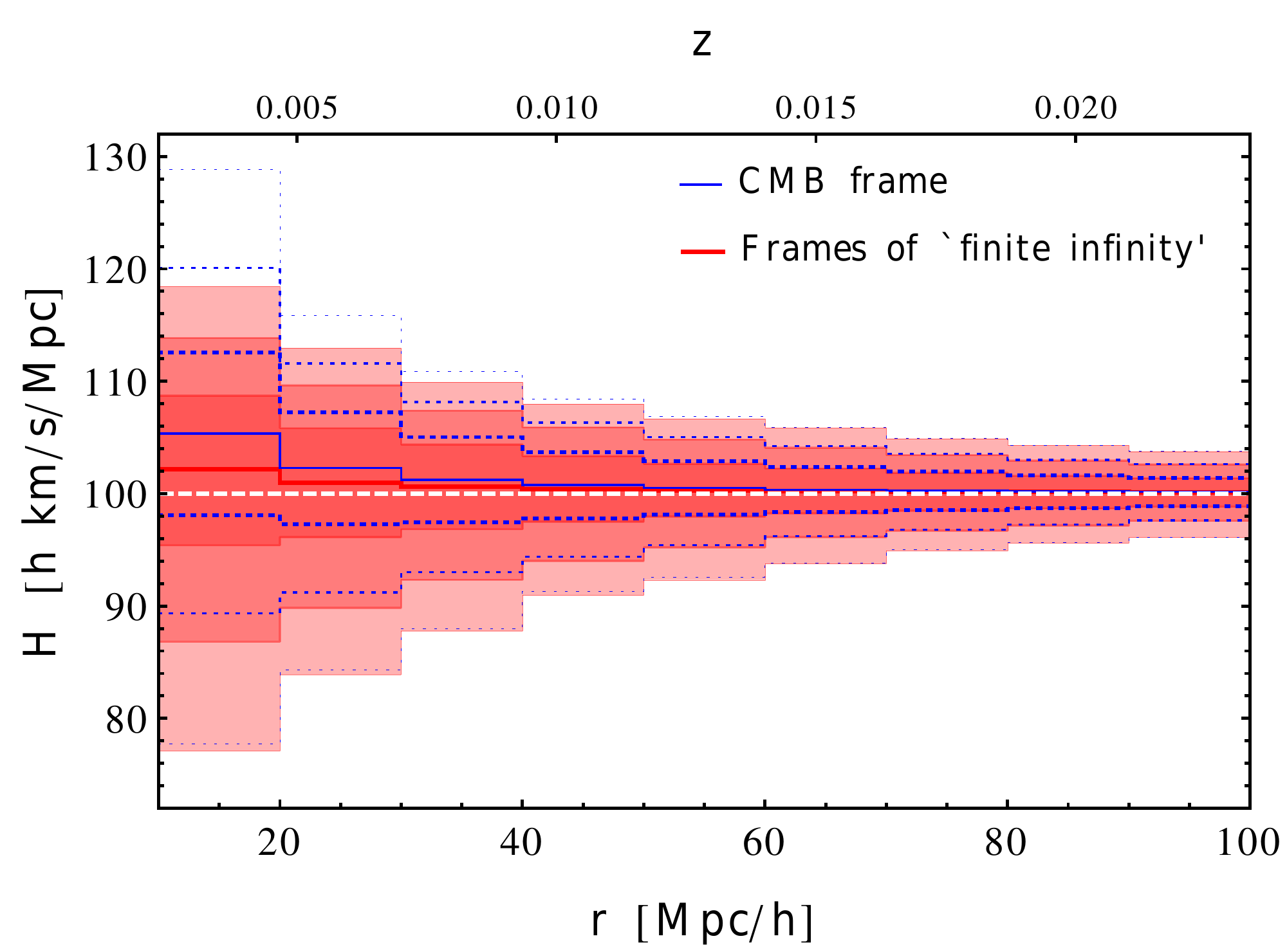}
 \caption{Hubble parameter as measured in the CMB frame (blue lines) and in frames of `finite infinity' (shaded red). The contours are 68.3\%, 95.4\% and 99.7\% confidence intervals while the solid lines are the means. Observers are located randomly. Note the small reduction in the Hubble variance (which is however still much bigger than in Fig.\ref{fig:Hvar}).}
   \label{fig:HvarFI}
\end{figure}

We examine this claim more closely by reversing the logic and calculating the Hubble parameter in the frames of `finite infinity' for each observer (see Fig.\ref{fig:HvarFI}). The Hubble parameter is indeed closer to its background value and has smaller variance compared to the CMB frame. Again, this is unsurprising as the bulk velocity of the `finite infinity' region is correlated with the bulk velocity of a bigger region. Hence, boosting into the FI frame on average reduces the dipole in the velocity flow and lowers the estimate for $H$. For random observers that are more likely to be in underdense regions with $H>H_0$, this has the effect of making $H$ closer to the background value. However, comparing the frame of minimum Hubble flow variation (Fig.\ref{fig:Hvar}) to the `finite infinity' frame (Fig.\ref{fig:HvarFI}), we note that the Hubble flow is much less uniform in the latter showing that further boosts are still required.

\section{Conclusions}
\label{sec:conclusion}

We have studied the properties of boost velocities of observers with respect to the CMB frame that make the spherically averaged Hubble flow converge the quickest to its background value. We place observers at different locations in a Hubble volume N-body simulation and find that the distribution of boost velocities they observe is near Gaussian with the amplitude dependent on the  location --- the larger the overdensity the larger the amplitude of the typical boost required. For observers in underdense regions, on average, the boosts that make the spherically averaged Hubble parameter converge fastest to the background value reduce at the same time the dipole structure of the Hubble flow. Based only on such local measurements of the velocity field the observers would choose such a frame as the cosmic rest frame given that it is closest to the na\"ive FLRW expectation. However for observers in overdense regions, on average, the boosts that make the Hubble parameter closer to the background value increase the dipole of the velocity field. We show that the boost velocity to the frame of most uniform flow is correlated with the bulk flow velocities and in particular with the group velocity of the `finite infinity' region, as was suggested in \citep{Wilthvar2}. The amplitude of the boost for our position \citep{Wilthvar2} in the universe is found \emph{not} to be in tension with the $\Lambda$CDM expectations. Note that the effects studied here are most pronounced in the mildly non-linear and non-linear regime, i.e. below the $z = 0.023$ threshold adopted \citep{riesshubble} in the determination of $H_0$ from SNe~Ia data. 

Additionally we re-derived the expression for the systematic offset of the Hubble parameter between different frames, noting that the dipole structure in spherical shells cannot be boosted away entirely. Our expression agrees with the measured difference between $H_\text{CMB}$ and $H_\text{LG}$ and resolves discrepancies in previous work.

\section{Acknowledgements}

DK thanks STFC UK for support and SS acknowledges a DNRF Niels Bohr Professorship. We thank Roya Mohayaee, Pavel Naselsky, and especially, David Wiltshire for many helpful discussions. We appreciate constructive comments by the anonymous referee.

\bibliographystyle{JHEP.bst}
\bibliography{refs_balt}

\end{document}